\documentclass[ reqno,11pt]{amsart}
\usepackage{graphicx,amsmath,amssymb,epsfig}
\usepackage{amsfonts}
\usepackage{rotating}
\usepackage{hyperref}

\hypersetup{
    colorlinks=true,
    urlcolor=blue,
    linkcolor=blue,
    citecolor=blue
}

\newtheorem{theorem}{Theorem}
\theoremstyle{plain}

\newtheorem{definition}{Definition}

\newtheorem{lemma}{Lemma}
\newtheorem{fact}{Fact}

\numberwithin{equation}{section}
\input{tcilatex}

\begin{document}
\title[Matching in Closed-Form]{Matching in Closed-Form:\\
Equilibrium, Identification, and Comparative Statics}
\author[Bojilov]{Raicho Bojilov{\small $^{\S }$}}
\author[Galichon]{Alfred Galichon\textit{$^{\dag }$}}
\date{Date: January 12, 2016. A preliminary version of this paper containing
the main results was first presented in September 2012 under the title
\textquotedblleft Closed-Form Formulas for Multivariate
Matching\textquotedblright . We would like to thank Nicholas Yannelis, the
Editor, and two anonymous referees, as well as Arnaud Dupuy, Bernard Salani%
\'{e} and seminar participants at CREST and the 2013 Search and Matching
Conference in Paris for helpful comments. Galichon's research has received
funding from the European Research Council under the European Union's
Seventh Framework Programme (FP7/2007-2013) / ERC grant agreements no
313699, and from FiME, Laboratoire de Finance des March\'{e}s de l'Energie.
Bojilov's work is supported by a grant of the French National Research
Agency (ANR), "Investissements d'Avenir" (ANR-11-IDEX-0003/Labex
Ecodec/ANR-11-LABX-0047).}

\begin{abstract}
This paper provides closed-form formulas for a multidimensional two-sided matching problem with transferable utility and heterogeneity in tastes. When the matching surplus is quadratic, the marginal distributions of the characteristics are normal, and when the heterogeneity in tastes is of the continuous logit type, as in Choo and Siow (2006), we show that the optimal matching distribution is also jointly normal and can be computed in closed form from the model primitives. Conversely, the quadratic surplus function can be identified from the optimal matching distribution, also in closed-form. The closed-form formulas make it computationally easy to solve problems with even a very large number of matches and allow for quantitative predictions about the evolution of the solution as the technology and the characteristics of the matching populations change.
\end{abstract}

\maketitle

\noindent


{\footnotesize \textbf{JEL codes}: C78, D61, C13.\vskip50pt }



\pagebreak

\section{Introduction}

Many economic problems involve markets with supply and demand restricted to
a unit of an indivisible good. Examples include occupational choice, task
and schedule assignment, sorting of CEOs to firms, venture capital
investment, consumer choice of heterogeneous goods, and the marriage market.
Models of discrete choice have been applied often to the empirical analysis
of such problems. While having important advantages, they generally do not
incorporate scarcity constraints and equilibrium price effects. Assignment
models provide an alternative framework which accommodates these issues. In
assignment models, the supply of goods of each type is fixed, and prices
adjust at equilibrium so that supply and demand clear. Yet, there have been
just a few general theoretical results in this environment beyond the proof
of existence, while econometric work has advanced only recently. The
pioneering work by Choo and Siow \cite{CS} proposes an econometric framework
for the estimation of assignment models based on the introduction of
logit-type unobserved heterogeneity in tastes. Their equilibrium approach
has a number of advantages: it incorporates unobserved heterogeneity, allows
for matching on many dimensions, and easily lends itself to nonparametric
identification. However, the equilibrium quantities in Choo and Siow's
setting are defined implicitly by a set of nonlinear equations, and
explicit, closed-form solutions, even under further assumptions on the
primitives of the model, have been missing so far. The present contribution
addresses this issue and also proposes a simple model with a complete
closed-form characterization of the equilibrium.

This paper considers an environment of one-to-one matching with transferable
utility when sorting occurs on multiple dimensions in the presence of
unobserved heterogeneity. Our goal is to develop a simple and easy to
estimate framework that can be used in both applied and empirical problems.
Thus, we study two related issues: on the one hand, solving for the optimal
matching distribution, given a surplus technology and the characteristics of
the matching populations; and, on the other hand, finding the surplus
technology, given the optimal matching and the characteristics of the two
sides of the market. We follow the setting of Choo and Siow \cite{CS} and
impose two additional assumptions: quadratic matching surplus and normality
on the distributions of the characteristics of the matching parties. These
assumptions play a key role in the derivation of our main result: they are
strong but not uncommon in the related literature. In particular, economists
have recognized the usefulness of the quadratic-normal setting and employed
it in models of hedonic pricing and matching since Tinbergen \cite{Tj}. In
addition, the introduction of randomness in the form of taste shocks has not
only some theoretical appeal, but it is also indispensable econometrically.
As Choo and Siow \cite{CS} discuss, such unobserved heterogeneity is crucial
for identification in matching models, even when the characteristics of the
matching parties are continuous.

We show that under the model assumptions both the identification and
equilibrium problems have explicit solutions in closed-form. In the
equilibrium problem, the optimal matching distribution is a multivariate
Gaussian distribution, and we provide explicit expressions for the relation
between the parameters of the optimal matching distribution and the
primitives of the model. Moreover, we derive closed-form expressions for
social welfare, equilibrium transfers, and the payoffs of market
participants. In the identification problem, we derive a closed-form
expression for the maximum-likelihood estimator of the surplus technology.
In contrast to estimation methods based on numerical methods, these
closed-form formulas make it computationally easy to solve problems that
involve a large number of matches and characterize the associated
equilibrium. They also allow for quantitative predictions about the sorting
that will occur and how it evolves as the complementarity on various
dimensions changes or as the distribution of characteristics in the
populations varies.

\textbf{Literature Review.} This paper is closely related to the literature
on identification of the matching function when the surplus is unobservable
(see \cite{Graham} for a good survey), following Choo and Siow \cite{CS}.
Fox \cite{F} proposes a maximum score estimator which relies on a rank-order
property. Galichon and Salani\'{e} \cite{GS} show that the social welfare in
this setting has a tractable formulation involving the entropy of the joint
distribution, from which identification can be deduced. They (\cite{GS}, 
\cite{GS2}) further introduce a parametric estimator of the surplus
function. Dupuy and Galichon \cite{DG} extend the model to the continuous
case and propose a decomposition of the surplus function into indices of
mutual attractiveness, in order to best approximate the matching patterns by
lower-dimensional models, and estimate the number of relevant dimensions on
which the sorting effectively occurs. In the case of matching with
non-transferable utility, Menzel \cite{Menzel} provides a characterization
of the equilibrium in a continuous logit framework, which, as we discuss in
the conclusion, could also be handled by our approach. Relative to these
papers, we derive a closed-form formula for the maximum likelihood estimator
in the normal-quadratic setting, and we present comparative statics that
investigate the sensitivity of the estimates to changes in the observed
data. Moreover, our results allow for easy counterfactual experiments.

A number of papers provide closed-form formulas for the equilibrium matching
problem with quadratic surplus and Gaussian distributions of the
populations, in the case when there is no unobserved heterogeneity.\ Early
examples include mathematically oriented papers by Olkin and Pukensheim \cite%
{OP}, as well as Dowson and Landau \cite{DL}. More recently, Lindenlaub \cite%
{L} and Machado \cite{Machado} work within the same setting to study
assortative matching and complementarities in multidimensional economic
settings. Our paper also connects with the literature on characterizing
matching equilibria and comparative statics in matching markets. This
literature includes, for example, contributions by Gretsky, Ostroy, and Zame 
\cite{GOZ} who analyze of the welfare properies of non-atomic assignment
models, and Ekeland Galichon, and Henry \cite{EGH} who investigate the
falsifiability of incompletely specified matching models. Moreover, Decker
et al. \cite{DLMS} study the equilibrium problem in Choo and Siow's discrete
setting and provide a number of comparative statics with a focus on the
impact of a change in the size of matching populations of given
characteristics. In relation to these works, we find closed-form solutions
to both the identification and equilibrium problems with heterogeneity. \ In
addition, we derive closed-form formulas for the optimal matching
distribution, social welfare, transfers, individual utility and firm
profits. Furthermore, we conduct a comparative statics exercise in a
continuous multidimensional setting to quantify how the characterization
varies with changes in the properties of the surplus technology and the
distributional characteristics of the populations.

Moreover, our work relates to the literature on hedonic pricing, which
includes \ the famous normal-quadratic model of Tinbergen \cite{Tj} and the
seminal contribution of Rosen \cite{Rs}. Brown and Rosen \cite{BR} suggest
the use of linear approximations to the equilibrium conditions for empirical
work and this approach underpins most of the following empirical studies.
Ekeland, Heckman, and Nesheim \cite{EHN} investigate the empirical content
of various hedonic pricing models, specifically models with quasi-linear
preferences and the normal-quadratic model. Chiappori, McCann, and Nesheim 
\cite{CMN} establish the link between models of hedonic pricing and
matching: they\ show that models of hedonic pricing with quasi-linear
preferences are equivalent to models of stable matching with transferable
utility. Furthermore stable matching assignments in the latter correspond to
equilibrium assignments in the former and the same equivalence applies for
prices. Finally, Ekeland \cite{E} investigates further the existence,
uniqueness, and efficiency of hedonic pricing models. Given the equivalence
in the environments, our work also contributes to this literature by
presenting a multidimensional model with closed-form solutions. In contrast,
the famous linear-quadratic-normal model in Tinbergen \cite{Tj} yields
closed-fom solutions only in the special case when the surplus technology
and the variances of the characteristics of the two sides in the market are
diagonal. As Ekeland, Heckman, and Nesheim \cite{EHN} \ point out (p. S70)
\textquotedblleft effectively, this is a scalar case in which each attribute
is priced separately.\textquotedblright 

The setting in the paper shares common features with several other
literatures. In the case without unobserved heterogeneity, the solution to
this equilibrium problem is well-known in the applied mathematics literature
and solved e.g. in \cite{DL}. Our mathematical contribution here is to move
beyond this benchmark and solve the problem with logit heterogeneity within
a framework of clear economic interpretation. The quadratic-normal setting
is used also in many-to-many matching, for example Gomes and Pavan \cite{GP}%
, and coordination games of incomplete information, as in Angeletos and
Pavan \cite{AP}. In addition, logit-type heterogeneity is a key component of
most choice models. Our use of logit heterogeneity in a continuous choice
setting relates to the spatial choice model of Ben-Akiva and Watanatada \cite%
{BW}. \ Relatedly, Dagsvik \cite{Dagsvik} provides a more general analysis
of continuous choice, max-stable processes, and the independence of
irrelevant alternative.

Finally, there is a vast panel of applications of assignment models. In the
literature on industrial organization, Fox and Kim \cite{FK} use an
extension of Choo and Siow's model to study the formation of supply chains.
In the corporate governance literature, Gabaix and Landier \cite{GL} and
Tervi\"{o} \cite{T} applied an assignment model to explain the sorting of
CEOs and firms and the determinants of the CEO compensation. In the marriage
literature, Chiappori, Salani\'{e} and Weiss \cite{CSW} used a
heteroskedastic version of Choo and Siow's model to estimate the changes in
the returns to education on the US marriage market; Chiappori, Oreffice and
Quintana-Domeque \cite{COQ} study the interplay between the sorting on
anthropomorphic and on socioeconomic characteristics.\ The context of our
model is closest to Hsieh, Hurst, Jones, and Klenow \cite{HHJK} who use a
variation of the Roy model to study discrimination in the US. This list is
necessarily very incomplete.

\textbf{Organization of the paper}. The rest of the paper is organized as
follows. Section~\ref{sec:Model} presents the model and states the optimal
matching problem. Section~\ref{sec:formulas} starts with the set of
conditions that link the solution of the matching problem with the
parameters of the social surplus and then derives closed-form expression for
the optimal matching, social surplus, wage transfers, individual utilities
and profits. Then, it presents the closed-form solution to the
identification problem and concludes with asymptotic analysis and
comparative statics of the closed-form solutions. Section~\ref%
{sec:conclusion} concludes.

\textbf{Notation}. Throughout the paper, we denote the transpose of matrix $%
M $\ as $M^{\ast }$\ and the inverse of this transpose as $M^{-\ast }.$\
Similarly, the inverse of the square-root of matrix $M,$ $M^{\frac{1}{2}},$\
is denoted $M^{-\frac{1}{2}}.$\ Let the generalized inverse\ (see~\cite%
{BenIsrael}) of a non-square matrix $M$ be denoted $M^{+}$and its transpose, 
$M^{+\ast }.$ All distributions are centered at $0$.\ The random variables
are denoted with capital letters, while the corresponding sample points are
denoted with small letter.

\bigskip

\section{Model\label{sec:Model}}

We consider a market for heterogeneous goods with demand and supply limited
to unity, in which transfers enter additively in the utility functions. We
assume away market frictions and focus on long-term outcomes and contexts in
which frictions play a relatively small role. The main advantage of applying
one-to-one matching with transferable utility (hedonic pricing) to the
analysis of such environments consists in the possibility to study the
interdependence of sorting (assignment) due to equilibrium price effects.
Due to its tractability, this setting has been used numerous times to study
marriage, CEO pay, venture capital, occupational choice and wage
differentials, etc. In contexts in which market frictions are of primary
interest, Shimer and Smith \cite{SS} provide an alternative approach that
extends the transferable utility framework to environments with both search
frictions and matching. Their focus is on the conditions under which
assortative matching may be preserved in this more complex environment.
While appealing because of its realism, this model has proven to be
operationally challenging in applied and empirical work, particularly in
multidimensional environments. Still further, the classical models of search
by inspection and by experience provide an alternative when there are search
frictions but interdependence of sorting decisions is not of importance.

We present the theoretical model in the context of labor matching between a
population of firms and a population of workers. We adopt the setting of
Dupuy and Galichon \cite{DG}, extending Choo and Siow \cite{CS} to the case
when the characteristics of the matching populations are continuous. As it
is standard in the search and matching literature, we interpret each firm as
a unique job position or task. A worker can be employed by only one firm and
a firm can employ only one worker. The populations are assumed to be of
equal size, and we assume that it is always better to be matched than to
remain alone and generate no surplus. Specifically, the focus is on markets
in which remaining unmatched is empirically of little relevance. Examples
include the markets for CEOs, sport managers and trainers, or specialists
like accountants, lawyers, and PR agents. The model also covers some cases
of one-to-many matching in which the profit of each firm is additive in the
characteristics of the workers, i.e. there are no externalities between the
assignment of individuals to the same firm.

\subsection{Environment}

Each worker has a vector of observable characteristics $x$ known to the
econometrician\textbf{, }where $x\in R^{m}$. Similarly, vector $y$
summarizes the observable characteristics of firms, where $y\in R^{n}.$ \ We
assume that the surplus enjoyed by worker $x$ choosing occupation $y$ is the
sum of three terms: one reflecting her intrinsic taste for occupation $y$,
as predicted by her observable type $x$; one reflecting the amount of net
monetary compensation; and one reflecting unobserved heterogeneity in tastes
for occupation $y$. In general, it is sufficient that the random taste
shocks have a known distribution conditional on characteristics $x$. We
follow the commonly adopted assumption in the literature and consider the
special case when the taste shocks are independent of $x.$ The formal
assumptions of the model are presented below.

\bigskip

\textbf{Assumption 1: Surplus.} The surplus of a worker with characteristics 
$x$ from matching with a firm with characteristics $y$ \ is%
\begin{equation}
x^{\prime }By+\tau \left( x,y\right) +\chi \left( y\right)  \label{workerut}
\end{equation}%
where $B$\ is a $m\times n$\ real matrix, $x^{\prime }By$\ is the
nonpecuniary amenity of career with firm $y$, $\tau \left( x,y\right) $ is
the monetary transfer or compensation received by the worker from the firm,
and $\chi \left( y\right) $ is a random utility process drawn by each
individual worker and whose distribution is characterized in Assumption 3
below. Similarly, the surplus of a firm with characteristics $y$ from
matching with a worker of characteristics $x$ is%
\begin{equation}
x^{\prime }\Gamma y-\tau \left( x,y\right) +\xi \left( x\right)
\label{firmpr}
\end{equation}%
where $\Gamma $\ is a $m\times n$\ real matrix, $x^{\prime }\Gamma y$\ is
the economic value created by worker $x$ at firm $y$, and $\xi \left(
x\right) $ is a random productivity process whose distribution is also
characterized in Assumption 3 below.

\bigskip

Assumption 1 specifies the form of heterogeneity that we investigate, namely
heterogeneity in the preferences over the observable characteristics of
one's partner. As a result of this assumption, the joint matching surplus is
separable in the sense that 
\begin{equation*}
\tilde{\Phi}=x^{\prime }Ay+\chi \left( y\right) +\xi \left( x\right)
\end{equation*}%
where the $m\times n$ real matrix $A,$ $A=B+\Gamma ,$ is called the \emph{%
affinity matrix}. The first term in the formula for the joint surplus is the
observable surplus. The second term allows for unobserved variation in the
preference of workers for observed firm characteristics, and the third
allows for unobserved variation in the preferences of firms over the
observed characteristics of workers. In other words, we rule out an
idiosyncratic term that represents the interaction of unobserved firm and
worker characteristics.

Thus, Assumption 1 follows Choo and Siow \cite{CS} in the way it specifies
the firm and worker surplus from matching, except for the specific
functional form of the surplus imposed in our setting. The quadratic
specification is the simplest nontrivial form of the joint surplus yielding
complementarities between any pair of firm and worker characteristics. This
specification provides an intuitive and meaningful interpretation of the
interaction between the characteristics of firms and workers: namely, $%
A_{ij} $ is simply the strength of the complementarity (positive or
negative) between the $i$-th characteristics of the firm and the $j$-th
characteristics of the worker. For this reason, it has been widely used in
empirical work and applied models since Tinbergen \cite{Tj}. Finally, note
that $x$ and $y$ can be of different dimension.

'

\textbf{Assumption 2: Unobservable Heterogeneity. }The function $\chi \left(
.\right) $ is modelled as an Extreme-value stochastic process 
\begin{equation*}
\chi \left( y\right) =\max_{k}\left\{ -\lambda \left( y_{k}-y\right) +\sigma
_{1}\chi _{k}^{i}\right\}
\end{equation*}%
where $\left( y_{k},\chi _{k}^{i}\right) $ are the points of a Poisson
process on $\mathbb{R}^{n}\times \mathbb{R}$ of intensity $dy\times e^{-\chi
}d\chi $ for worker\textbf{\ }$i$, and $\lambda \left( z\right) =0$ if $z=0$%
, $\lambda \left( z\right) =+\infty $ otherwise. Similarly, 
\begin{equation*}
\xi \left( x\right) =\max_{l}\left\{ -\lambda \left( x_{l}-x\right) +\sigma
_{2}\xi _{l}^{j}\right\}
\end{equation*}%
where $\left( x_{l},\xi _{l}^{j}\right) $ are the points of a Poisson
process on $\mathbb{R}^{m}\times \mathbb{R}$ of intensity $dx\times e^{-\xi
}d\xi $ for firm $j$.

\bigskip

Assumption 2 implies that each worker and each firm draws an infinite, but
discrete number of \textquotedblleft acquaintances\textquotedblright\ from
the opposite side of the market, along with a random surplus shock. In
essence, this specification allows for the extension of the convenient logit
probability function to the analysis of continuous choice problems. Let $%
\sigma $\ be the sum of the scale parameters $\sigma _{1}$\ and $\sigma _{2}$%
\ defined in Assumption 2. It captures the total amount of heterogeneity and
will play an important role in the sequel. Further details on the continuous
logit model can be found in \cite{BW}.

\bigskip

\textbf{Assumption 3: Observable Heterogeneity}. We assume that $X$ has a
Gaussian distribution $P=N\left( 0,\Sigma _{X}\right) $ and, similarly, that 
$Y$ has distribution $Q=N\left( 0,\Sigma _{Y}\right) $.

\bigskip

Assumption 3 is arguably the strongest restriction of the model. It imposes
normal marginal distributions for the characteristics of the matching
populations. The theoretical literature has long recognized the usefulness
of the quadratic-normal setting, starting with Tinbergen \cite{Tj}.
Following our discussion of comparative statics below, we propose an
overidentification test that can be used to determine whether the model
assumption, and in particular the quadratic-nomal setting, are statistically
acceptable. Empirically, the normality assumption is appropriate in many
cases, while this is not the case in others due to the presence of discrete
characteristics, skewness, fat tails, etc. For example, researchers have
often found that the normal distribution provides a good approximation to
the actual distribution of IQ tests, test of non-cognitive skills, height
and weight. Moreover, most of the applied and empirical literature following
the seminal contribution of Rosen \cite{Rs} is also based on the
quadratic-normal setting. As discussed in Brown and Rosen \cite{BR},
researchers have interpreted the equilibrium conditions associated with the
setting as linear approximations to the actual equilibrium conditions. The
same quadratic-normal setting plays the role of a benchmark case in the
identification analysis of hedonic models in Ekeland, Heckman, and Nesheim 
\cite{EHN}. \ Finally, quadratic-Gaussian models have a long tradition also
in coordination games under incomplete information, for example Angeletos
and Pavan \cite{AP}, and in the analysis of many-to-many matching with
multidimensional characteristics, including Gomes and Pavan \cite{GP}, and
the references therein.

'

\textbf{Assumption 4: Information.} The firms and workers know $x$, $y,$\
and the realized taste shocks at the time they decide with whom to match,
but the econometrician observes only $x$\ and $y$.

This assumption defines the information structure of the problem: agents
have full information, the econometrician does not observe preference shocks.

\subsection{Equilibrium Conditions}

Given Assumptions~1 to 4, a matching assignment is a distribution $\pi $\
over the characteristics of firms and workers $\left( X,Y\right) $\ with
marginals for $X$\ and $Y$\ equal to $P$\ and $Q$\ respectively. Let $%
\mathcal{M}\left( P,Q\right) $\ be the set of matching assignments, i.e.
distributions over $\left( X,Y\right) ,$\ such that $X\sim P$, and $Y\sim Q$%
. We start by reviewing the equilibrium conditions that establish a link
between the primitives of the model, particularly the surplus technology,
and the optimal matching, denoted $\pi _{XY}$. As usual, workers maximize
utility by solving the following problem:%
\begin{equation*}
\max_{y}\text{ }x^{\prime }By+\tau \left( x,y\right) +\chi \left( y\right)
\end{equation*}%
From the properties of the Generalized Extreme Value (GEV) distribution, 
\begin{equation}
\log \pi _{XY}\left( x,y\right) =\frac{x^{\prime }By+\tau \left( x,y\right)
-a\left( x\right) }{\sigma _{1}}  \label{Worker_optimality}
\end{equation}%
where $a\left( x\right) $\ is a normalization term that depends only on $x$%
\textbf{\ }%
\begin{equation*}
a\left( x\right) =-\sigma _{1}\log \frac{f\left( x\right) }{\int \exp \left[ 
\frac{x^{\prime }By+\tau \left( x,y\right) }{\sigma _{1}}\right] dy}
\end{equation*}%
and $f\left( x\right) $\ is the probability density function associated with
characteristics $X$.\textbf{\ }Similarly, one obtains another condition that
links the optimal matching distribution and the technology of the firm $%
x^{\prime }\Gamma y:$ 
\begin{equation}
\log \pi _{XY}\left( x,y\right) =\frac{x^{\prime }\Gamma y-\tau \left(
x,y\right) -b\left( y\right) }{\sigma _{2}}  \label{Firm_optimality}
\end{equation}%
where 
\begin{equation*}
b\left( y\right) =-\sigma _{2}\log \frac{g\left( y\right) }{\int \exp \left[ 
\frac{x^{\prime }\Gamma y-\tau \left( x,y\right) }{\sigma _{2}}\right] dx}
\end{equation*}%
and $g\left( y\right) $\ is the probability density function associated with
characteristics $Y$. Equilibrium in the market requires that supply equals
demand and, therefore, the probabilities in equations (\ref%
{Worker_optimality}) and (\ref{Firm_optimality}) must be consistent. Using
this equilibrium requirement, we solve for $\tau \left( x,y\right) $ and $%
\pi _{XY}\left( x,y\right) $ by combining (\ref{Worker_optimality}) and (\ref%
{Firm_optimality}) to obtain:%
\begin{eqnarray}
\log \pi _{XY}\left( x,y\right) &=&\frac{x^{\prime }Ay-a\left( x\right)
-b\left( y\right) }{\sigma }  \label{EqProba} \\
\tau \left( x,y\right) &=&\frac{\sigma _{1}\left( x^{\prime }\Gamma
y-b\left( y\right) \right) -\sigma _{2}\left( x^{\prime }By-a\left( x\right)
\right) }{\sigma }  \label{EqTransfers}
\end{eqnarray}%
where $\sigma =\sigma _{1}+\sigma _{2}$. To complete the conditions that
characterize the equilibrium, we define the feasibility constraints: $\pi
_{XY}\in \mathcal{M}\left( P,Q\right) .$

These equilibrium conditions coincide with the conditions that characterize
the solution to the associated social welfare problem. We recall the
characterization of the equilibrium given by Dupuy and Galichon \cite{DG},
who show how the problem of estimating the equilibrium allocation, or
optimal matching, can be solved by estimating the corresponding social
welfare problem of the planner. Theorem 1 in~\cite{DG} implies that:

(i) At equilibrium, the optimal matching $\pi _{XY}\in \mathcal{M}\left(
P,Q\right) $\ is a maximizer of the social welfare%
\begin{equation}
\mathcal{W}\left( A\right) =\sup_{\pi \in \mathcal{M}\left( P,Q\right)
}\left( \mathbb{E}_{\pi }\left[ X^{\prime }AY\right] -\sigma \mathcal{E}%
\left( \pi \right) \right)  \label{OptPi}
\end{equation}%
where\textit{\ }%
\begin{eqnarray*}
\mathcal{E}\left( \pi \right) &=&\mathbb{E}_{\pi }\left[ \ln \pi \left(
X,Y\right) \right] \text{ if }\pi \text{ is absolutely continuous} \\
&=&+\infty \text{ otherwise.}
\end{eqnarray*}

(ii) $\pi _{XY}$\ is solution of (\ref{OptPi}) if and only if it is a
solution of%
\begin{equation}
\left\{ 
\begin{array}{c}
A_{ij}=\sigma \frac{\partial \log \pi _{Y|X}\left( y|x\right) }{\partial
x_{i}\partial y_{j}} \\ 
\pi _{XY}\in \mathcal{M}\left( P,Q\right)%
\end{array}%
\right. .  \label{condPi}
\end{equation}

(iii) for any $\sigma >0$, the solution of (\ref{condPi}) exists and is
unique.

\bigskip

In particular, part (ii) establishes the equivalence. Differentiating
condition (\ref{EqProba}), we obtain the first part of equilibrium
conditions (\ref{condPi}), while the second part is just a restatement of
the feasibility constraints. The equivalence of the matching problem to a
linear optimization problem helps provide an expression for the social gain
from matching, defined in part (i) above. This result can be seen as an
extension of the Monge-Kantorovich theory (see Chapter 2 of \cite{Vi}). It
implies that the introduction of unobserved heterogeneity naturally leads to
the classical matching problem with an additional information term that
attracts the optimal solution toward a random matching: the entropy of the
joint distribution $\mathbb{E}_{\pi _{XY}}\left[ \ln \pi _{XY}\left(
X,Y\right) \right] .$ If $\sigma $ is large, then optimality requires
minimizing the mutual information which happens when firms and workers are
matched randomly to each other. On the other hand, if $\sigma $ is small,
optimality requires maximizing the observable surplus. Uniqueness of the
solution is a more general result in \cite{DG}, but in our specific case it
can be simply verified after solving the problem.

Finally, the optimality conditions for the worker (\ref{Worker_optimality})
and the equilibrium condition (\ref{EqProba}) yield an expression for the
deterministic part of the utility of a worker of characteristics $x$\
matching with a firm of characteristics $y$: 
\begin{equation*}
U\left( x,y\right) =\frac{\sigma _{1}x^{\prime }Ay-\sigma _{1}b\left(
y\right) +\sigma _{2}a\left( x\right) }{\sigma }
\end{equation*}%
Similarly, we use (\ref{Firm_optimality}) and (\ref{EqProba}) to obtain the
profits of a firm with characteristics $y$\ matching with a worker with
characteristics $x$: 
\begin{equation*}
\Pi \left( x,y\right) =\frac{\sigma _{2}x^{\prime }Ay-\sigma _{2}a\left(
x\right) +\sigma _{1}b\left( y\right) }{\sigma }
\end{equation*}%
The expressions $a\left( x\right) $\ and $b\left( y\right) $\ correspond to
the Lagrange multipliers associated with the scarcity constraints for
workers of characteristics $x$\ and for firms of characteristics $y.$\
Consequently, the contribution of $a\left( x\right) $\ to individual utility
captures the intuition that the share of the surplus that goes to a
particular worker increases in the scarcity of her characteristics$.$\
Interestingly, the utility also increases in the scale of the taste shocks
over the firm characteristics. A similar interpretation also holds for
profits.

\section{Closed-Form Formulas\label{sec:formulas}}

This section contains our main results. We start by discussing how the
quadratic setting and the distributional assumptions give rise to a
tractable condition that allows us to find a closed-form formula for the
optimal matching in terms of the affinity matrix $A,$ and the parameters
that describe the distributions of $\ X$ and $Y.$ Then, we use the same
condition to recover the expression of the surplus, namely the affinity
matrix $A,$ in terms of the observed data. Denote%
\begin{equation}
\Sigma _{XY}=\left( \mathbb{E}_{\pi _{XY}}\left[ X_{i}Y_{j}\right] \right)
_{ij}=\mathbb{E}_{\pi _{XY}}\left[ XY^{\prime }\right]  \label{crosscov}
\end{equation}%
to be the \emph{cross-covariance matrix} computed at the optimal $\pi _{XY}$
solution of (\ref{condPi}). We will show that under assumptions 1 to 4 the
optimal solution is normal and that $\left( \Sigma _{X},\Sigma _{Y},\Sigma
_{XY}\right) $ completely parameterize the distribution $\pi _{XY}$.

\bigskip

We consider two related problems. The equilibrium computation problem
requires, given a surplus technology, to determine the optimal matching $\pi
_{XY}$. Finding out how to do this in a tractable way, possibly in closed
form, allows to make quantitative predictions about the sorting that will
occur on the market and to derive comparative statics. In contrast, the
identification problem starts with an observed matching that is assumed to
be stable (or equivalently, optimal) and the goal is to determine the
underlying surplus technology. This problem is the inverse problem of the
former and is of primary interest to the econometrician. The conditions (\ref%
{condPi}), in particular the first one, play a crucial role in our analysis
because they establish a link between the cross-covariance matrix and the
affinity matrix. Thus, the equilibrium computation problem involves solving
the first-order conditions (\ref{condPi}) for the optimal matching and $%
\Sigma _{XY},$ while the identification problem involves solving the same
conditions but for $A$ given $\Sigma _{XY}.$ Closed-form formulas shall be
provided for these two problems.

\subsection{Equilibrium Computation Problem}

The following result provides an explicit solution to the equilibrium
problem. It requires some educated `guesswork' and consists of two parts. In
the first part, we propose a multivariate normal distribution as a candidate
for the solution, and then we verify that the candidate satisfies the
equilibrium conditions (\ref{condPi}). From above, we know that this
solution is unique. To the interested readers, we reveal how we arrive at
our candidate for the solution in the second part of the proof. Namely, we
start with the first condition in (\ref{condPi}) and note that in the case
of multivariate normal distributions it reduces to a quadratic matrix
equation in $\Sigma _{XY},$\ since the cross-derivative of the optimal
matching is constant. Most such equations do not have closed-form solutions
but we redefine variables in a way that allows us to solve for the unknown $%
\Sigma _{XY}$\ in closed-form. While members of the exponential family
satisfy some of the equilibrium conditions, we find that it is only the
multivariate normal distribution that satisfy all of them.

\begin{theorem}
\label{thm:optMatching}Assume $\sigma >0$ and suppose that Assumptions 1 to
4 hold. Let $\left( X,Y\right) \sim \pi _{XY}$ be the solution of (\ref%
{OptPi}) that satisfies conditions (\ref{condPi}). Then:

(i) The relation between $X$ and $Y$ takes the following form%
\begin{equation}
Y=TX+\epsilon  \label{regXY}
\end{equation}%
where $\epsilon \sim N\left( 0,\Sigma _{Y|X}\right) $ is a random vector
independent from $X$, and matrices $T$ and $\Sigma _{Y|X}$ are given by%
\begin{eqnarray}
T &=&\Delta \left( \Delta A^{\ast }\Sigma _{X}A\Delta \right) ^{-1/2}\Delta
A^{\ast }-\frac{\sigma }{2}A^{+}\Sigma _{X}^{-1}  \label{ExprT} \\
\Sigma _{Y|X} &=&\sigma \Delta \left( \Delta A^{\ast }\Sigma _{X}A\Delta
\right) ^{-1/2}\Delta -\frac{\sigma ^{2}}{2}A^{+}\Sigma _{X}^{-1}A^{+\ast }
\label{ExprV}
\end{eqnarray}%
where matrix $\Delta $ is defined as%
\begin{equation*}
\Delta =\left( \frac{\sigma ^{2}}{4}A^{+}\Sigma _{X}^{-1}A^{+\ast }+\Sigma
_{Y}\right) ^{1/2}.
\end{equation*}

(ii) The optimal matching $\pi _{XY}$ is the Gaussian distribution $N\left(
0,\Sigma \right) $ where 
\begin{equation}
\Sigma =%
\begin{pmatrix}
\Sigma _{X} & \Sigma _{XY} \\ 
\Sigma _{XY}^{^{\ast }} & \Sigma _{Y}%
\end{pmatrix}%
,  \label{Sigma}
\end{equation}%
and the cross-covariance matrix of $X$ and $Y$, namely $\Sigma _{XY}=\mathbb{%
E}_{\pi _{XY}}\left[ XY^{\prime }\right] $ is given by%
\begin{equation}
\Sigma _{XY}=\Sigma _{X}A\Delta \left( \Delta A^{\ast }\Sigma _{X}A\Delta
\right) ^{-1/2}\Delta -\frac{\sigma }{2}A^{+\ast }.  \label{SigXY}
\end{equation}
\end{theorem}

\begin{proof}
Under Assumptions 1-4, and as recalled in Section~\ref{sec:Model} the
solution to the matching assignment problem is unique and characterized by
conditions (\ref{condPi}). Hence it is sufficient just to verify that taking 
$\pi _{XY}$ as the p.d.f. of the $N\left( 0,\Sigma \right) $ distribution
where $\Sigma $ satisfies (\ref{Sigma}) and (\ref{SigXY}) satisfies
conditions (\ref{condPi}) characterizing the optimal assignment.

\bigskip

\textbf{Verification of optimality. }Let $\pi _{XY}$ be the distribution of $%
\left( X,Y\right) $ where $X\sim N\left( 0,\Sigma _{X}\right) $ and the
distribution of $Y$ conditional on $X$ is given by (\ref{regXY}), for $T$
and $\Sigma _{Y|X}$ given by (\ref{ExprT}) and (\ref{ExprV}). We verify that 
$\pi _{XY}$ satisfies the two conditions of characterization (\ref{condPi}).
Indeed, one has%
\begin{equation*}
\partial _{xy}^{2}\log \pi _{Y|X}\left( y|x\right) =T^{\ast }\Sigma
_{Y|X}^{-1}=\frac{A}{\sigma }
\end{equation*}%
and%
\begin{eqnarray*}
var\left( Y\right) &=&T\Sigma _{X}T^{\ast }+\Sigma _{Y|X} \\
&=&\left( \Delta \left( \Delta A^{\ast }\Sigma _{X}A\Delta \right)
^{-1/2}\Delta A^{\ast }-\frac{\sigma }{2}A^{+}\Sigma _{X}^{-1}\right) \Sigma
_{X}T^{\ast }+\Sigma _{Y|X} \\
&=&\Delta ^{2}-\sigma \Delta \left( \Delta A^{\ast }\Sigma _{X}A\Delta
\right) ^{-1/2}\Delta +\frac{\sigma ^{2}}{4}A^{+}\Sigma _{X}^{-1}A^{+\ast
}+\Sigma _{Y|X} \\
&=&\Delta ^{2}-\frac{\sigma ^{2}}{4}A^{+}\Sigma _{X}^{-1}A^{+\ast }=\Sigma
_{Y}.
\end{eqnarray*}%
Hence, condition (\ref{condPi}) is verified and $\pi _{XY}$ is optimal for (%
\ref{OptPi}), QED.

Although this proof by verification is sufficient from a pure mathematical
point of view, it is not very didactic as it is not informative about how
the formula for $\Sigma _{XY}$ was obtained. \ For the convenience of the
reader, we shall now flesh out how to solve for $\Sigma _{XY}$.

\bigskip

\textbf{Solving for }$\Sigma _{XY}$. In this paragraph, we look for a
necessary condition on $\Sigma $ so that $N\left( 0,\Sigma \right) $ is a
solution of (\ref{OptPi}). The derivation made before Theorem \ref%
{thm:MatchingEstimator} implies the following relation to be inverted
between $A$ and $\Sigma _{XY}$ 
\begin{equation*}
\frac{A}{\sigma }=\left( \Sigma _{Y}\left( \Sigma _{XY}\right) ^{+}\Sigma
_{X}-\Sigma _{XY}^{\ast }\right) ^{+}.
\end{equation*}

Some algebra leads to the following quadratic equation in $\Sigma _{XY}$%
\begin{equation}
0=\Sigma _{XY}^{\ast }\Sigma _{X}^{-1}\Sigma _{XY}+\sigma A^{+}\Sigma
_{X}^{-1}\Sigma _{XY}-\Sigma _{Y}  \label{Quadratic}
\end{equation}

Let 
\begin{eqnarray}
\Psi &=&\Sigma _{X}^{-1/2}\Sigma _{XY}  \label{ExprPsi} \\
D &=&\frac{\sigma }{2}A^{+}\Sigma _{X}^{-1/2}  \label{ExprB} \\
C &=&-\Sigma _{Y}  \label{ExprC}
\end{eqnarray}%
so that equation (\ref{Quadratic}) becomes%
\begin{equation}
\Psi ^{\ast }\Psi +2D\Psi +C=0  \label{quadeq2}
\end{equation}%
hence%
\begin{equation*}
2D\Psi =\Sigma _{Y}-\Sigma _{XY}^{\ast }\Sigma _{X}^{-1}\Sigma _{XY}
\end{equation*}%
is the variance-covariance matrix of the residual $\epsilon $ of the
regression of $Y$ on $X$%
\begin{equation}
\Sigma _{Y}=T\Sigma _{X}T^{\ast }+\Sigma _{Y|X}  \label{sigmaYformula}
\end{equation}%
and $T$ and $\Sigma _{Y|X}$ are given by expressions: 
\begin{equation}
T=\Sigma _{XY}^{\ast }\Sigma _{X}^{-1}  \label{FormT}
\end{equation}%
\begin{equation}
\Sigma _{Y|X}=\Sigma _{Y}-\Sigma _{XY}^{\ast }\Sigma _{X}^{-1}\Sigma _{XY}
\label{FormV}
\end{equation}

Hence, $D\Psi $\ is symmetric positive. Equation (\ref{quadeq2}) can be
rewritten as 
\begin{equation*}
\left( \Psi +D^{\ast }\right) ^{\ast }\left( \Psi +D^{\ast }\right)
=DD^{\ast }-C
\end{equation*}

Thus, we have for the solution that 
\begin{equation}
\Psi =U\Delta -D^{\ast }  \label{psieq}
\end{equation}%
where%
\begin{equation*}
\Delta =\left( DD^{\ast }-C\right) ^{1/2}=\left( \frac{\sigma ^{2}}{4}%
A^{+}\Sigma _{X}^{-1}A^{+\ast }+\Sigma _{Y}\right) ^{1/2}
\end{equation*}%
is a positive semi-definite matrix and $U$\ is an orthogonal matrix to be
defined. Note that $DD^{\ast }-C=DD^{\ast }+\Sigma _{Y}$\ is a positive
symmetric matrix so its square root exists and is uniquely defined.

We now determine $U$. Since $D\Psi $\ is symmetric positive, $DU\Delta
=D\Psi +DD^{\ast }$\ is symmetric positive, hence $\Delta ^{-1}DU=\Delta
^{-1}\left( DU\Delta \right) \Delta ^{-1}$\ as $\Delta $ is symmetric and
invertible. Let\textbf{\ }%
\begin{equation*}
\Lambda =\Delta ^{-1}D
\end{equation*}%
Since $\Lambda U$\ is symmetric positive, $\left( \Lambda U\right)
^{2}=\Lambda UU^{\ast }\Lambda ^{\ast }=\Lambda \Lambda ^{\ast }$, which
implies that $\Lambda U=\left( \Lambda \Lambda ^{\ast }\right) ^{1/2}$.
Hence $U$\ is determined by%
\begin{equation*}
U=\Lambda ^{-1}\left( \Lambda \Lambda ^{\ast }\right) ^{1/2}=D^{-1}\Delta
\left( \Delta ^{-1}DD^{\ast }\Delta ^{-1}\right) ^{1/2}.
\end{equation*}

Substituting in (\ref{psieq}) yields%
\begin{equation*}
\Psi =U\Delta -D^{\ast }=D^{-1}\Delta \left( \Delta ^{-1}DD^{\ast }\Delta
^{-1}\right) ^{1/2}\Delta -D^{\ast }
\end{equation*}%
and replacing $D,\Sigma _{XY}$, $T$\ and $\Sigma _{Y|X}$\ by their
respective expressions (\ref{ExprB}), (\ref{ExprPsi}), (\ref{FormT}) and (%
\ref{FormV}) gives%
\begin{eqnarray*}
\Psi &=&\Sigma _{X}^{1/2}A\Delta \left( \Delta ^{-1}\left( A^{+}\Sigma
_{X}^{-1}A^{+\ast }\right) \Delta ^{-1}\right) ^{1/2}\Delta -\frac{1}{2}%
\Sigma _{X}^{-1/2}A^{+\ast } \\
\Sigma _{XY} &=&\Sigma _{X}A\Delta \left( \Delta A^{\ast }\Sigma _{X}A\Delta
\right) ^{-1/2}\Delta -\frac{\sigma }{2}A^{+\ast } \\
T &=&\Delta \left( \Delta A^{\ast }\Sigma _{X}A\Delta \right) ^{-1/2}\Delta
A^{\ast }-\frac{\sigma }{2}A^{+}\Sigma _{X}^{-1} \\
\Sigma _{Y|X} &=&\sigma \Delta \left( \Delta A^{\ast }\Sigma _{X}A\Delta
\right) ^{-1/2}\Delta -\frac{\sigma ^{2}}{2}A^{+}\Sigma _{X}^{-1}A^{+\ast }.
\end{eqnarray*}
\end{proof}

\bigskip

Hence it turns out that the optimal matching solution $\pi _{XY}$ is jointly
normal too, a fact that was not a priori obvious, as $\pi _{XY}$ may have
normal marginal distributions for $X$ and $Y$ without necessarily being
jointly normal.

In what follows, we provide a characterization of the solution. We first
consider how the solution evolves both when the unobserved heterogeneity
declines to zero and when it increases to infinity. The following lemma
provides these limiting results, recovering well-known results in the
optimal transportation literature (see e.g. \cite{DL}).

\begin{lemma}
\label{lemmaLimits}\textit{Suppose that Assumptions 1 to 4 hold, and let }$%
\Sigma _{XY}$\textit{\ be the cross-covariance matrix under the optimal
matching. Then:}
\end{lemma}

\textbf{(i).} $\lim_{\sigma \rightarrow 0}\Sigma _{XY}=\Sigma _{X}A\Sigma
_{Y}^{1/2}\left( \Sigma _{Y}^{1/2}A^{\ast }\Sigma _{X}A\Sigma
_{Y}^{1/2}\right) ^{-1/2}\Sigma _{Y}^{1/2}$

\textbf{(ii).} $\lim_{\sigma \rightarrow \infty }\Sigma _{XY}=0.$

When the importance of unobserved variables increases, the optimal matching
ceases to depend on the observable characteristics $X$ and $Y,$ so in the
limit the cross-covariance matrix converges to zero. In the other limit
case, when there is no unobserved heterogeneity, the optimal matching
problem becomes a special case of Brenier's Theorem (see \cite{Vi}, Ch. 2)
and there exists an optimal matching map given by $Y=T\left( A^{\ast
}X\right) ,$\ where $T$\ is the gradient of a convex function which in our
setting has the economic interpretation of the derivative of the worker's
utility. Note that 
\begin{equation*}
\lim_{\sigma \rightarrow 0}T=T_{0}=\Sigma _{Y}^{1/2}\left( \Sigma
_{Y}^{1/2}A^{\ast }\Sigma _{X}A\Sigma _{Y}^{1/2}\right) ^{-1/2}\Sigma
_{Y}^{1/2}A^{\ast }
\end{equation*}%
and, consequently, $AT_{0}$\ is symmetric positive. Thus, the worker's
equilibrium utility$,$ when $\sigma =0,$\ is equal, up to an additive
constant, to $\frac{1}{2}x^{\prime }AT_{0}x.$\ 

In the likely presence of heterogeneity, however, there exists no such
optimal map which complicates the theoretical and econometric problems.
Nevertheless, expression (\ref{SigXY}) from Theorem \ref{thm:optMatching},
part i shows that it is still possible to establish a `regression-style'
relation between $X$\ and $Y$\ in our setting. It turns out that under the
optimal matching one can decompose $Y$\ into the sum of two independent
normally distributed terms: a linear combination of $X$\ (a projection of $Y$%
\ onto $X$) and an `error' whose distributional properties depend on the
primitives of the model, particularly the scale parameter of the unobserved
heterogeneity. This `regression-style' relation plays a crucial role in the
logic of the proofs of both Theorem \ref{thm:optMatching} and Theorem \ref%
{thm:MatchingEstimator}.

The closed-form solution for the optimal matching also allows us to derive a
closed-form expression for social welfare (up to an irrelevant integration
constant) under the optimal matching distribution: 
\begin{equation*}
\mathcal{W}\left( A\right) =Tr\left( A^{\ast }\Sigma _{XY}\right) -\sigma %
\left[ \frac{1}{2}\ln \left( \det \left( \Sigma _{X}\right) \det \left(
\Sigma _{Y}-\Sigma _{XY}^{\ast }\Sigma _{X}^{-1}\Sigma _{XY}\right) \right) %
\right]
\end{equation*}%
where $\Sigma _{XY}$\ is expressed as a function of $A$\ by~(\ref{SigXY}).
The first term is equal to the expected surplus from the observables and the
second one is equal to the entropy term. Using the equilibrium condition (%
\ref{EqProba}) in combination with Schur decomposition of the
variance-covariance matrix, we obtain the following expression for the
Lagrange multipliers$:$%
\begin{eqnarray*}
a\left( x\right) &=&\frac{\sigma }{2}x^{\prime }\Sigma _{X|Y}^{-1}x \\
b\left( y\right) &=&\frac{\sigma }{2}y^{\prime }\Sigma _{Y|X}^{-1}y
\end{eqnarray*}%
where $\Sigma _{X|Y}$\ and $\Sigma _{Y|X}$\ are the conditional variances of 
$x$\ and $y,$ given the optimal matching distribution defined in Theorem \ref%
{thm:optMatching}. Thus, the equilibrium transfer becomes 
\begin{equation*}
\tau \left( x,y\right) =\frac{1}{\sigma }\left( \sigma _{1}\left( x^{\prime
}\Gamma y-\frac{\sigma }{2}y^{\prime }\Sigma _{Y|X}^{-1}y\right) -\sigma
_{2}\left( x^{\prime }By-\frac{\sigma }{2}x^{\prime }\Sigma
_{X|Y}^{-1}x\right) \right)
\end{equation*}%
which has a very intuitive interpretation. The wage increases in the gains
generated by the firm from the match and the market scarcity of workers of
characteristics $x.$\ At the same time, it decreases in the worker from the
match and in the relative scarcity of firms of characteristics $y.$\ These
results imply that the expected utility at the equilibrium before the
realization of the taste shocks is:%
\begin{equation*}
U\left( x,y\right) =\frac{\sigma _{1}}{\sigma }x^{\prime }Ay-\frac{\sigma
_{1}}{2}y^{\prime }\Sigma _{Y|X}^{-1}y+\frac{\sigma _{2}}{2}x^{\prime
}\Sigma _{X|Y}^{-1}x
\end{equation*}%
while the associated expected profits are%
\begin{equation*}
\Pi \left( x,y\right) =\frac{\sigma _{2}}{\sigma }x^{\prime }Ay+\frac{\sigma
_{1}}{2}y^{\prime }\Sigma _{Y|X}^{-1}y-\frac{\sigma _{2}}{2}x^{\prime
}\Sigma _{X|Y}^{-1}x
\end{equation*}%
In other words, each side of the market receives a share of the joint
surplus that increases in the heterogeneity of the shocks over the
characteristics of the other side of the market. In addition, it obtains a
fraction of the equilibrium shadow price of its own characteristics and
`pays' a fraction of the other side's shadow price.\ Interestingly, the
division of the surplus as the unobserved heterogeneity vanishes depends on
the particular path of convergence of $\left( \sigma _{1},\sigma _{2}\right) 
$\ to $\left( 0,0\right) .$\ For example, when $\sigma _{1}=\sigma _{2}$,
which is a standard restriction in models of the marriage market, the
division coincides with the utility and profits presented after Lemma \ref%
{lemmaLimits}.

\subsection{Identification Problem}

The identification problem is simpler. From Theorem \ref{thm:optMatching}, $%
\left( X,Y\right) $ is jointly normal, so one may regress $Y$\ on $X$\ and
write%
\begin{equation}
Y=TX+\epsilon  \label{regXYbis}
\end{equation}%
where $\epsilon \sim N\left( 0,\Sigma _{Y|X}\right) $\ is independent from $%
X $. Consequently, we can express $\Sigma _{Y}=\mathbb{E}\left[ YY^{\prime }%
\right] $ and $\Sigma _{XY}^{\ast }=\mathbb{E}\left[ YX^{\prime }\right] $
as\ 
\begin{align*}
\Sigma _{Y}& =T\Sigma _{X}T^{\ast }+\Sigma _{Y|X} \\
\Sigma _{XY}^{\ast }& =T\Sigma _{X}
\end{align*}%
and, solving for $T$ and $\Sigma _{Y|X},$\ we arrive again at expressions (%
\ref{FormT}) and (\ref{FormV}):%
\begin{eqnarray*}
T &=&\Sigma _{XY}^{\ast }\Sigma _{X}^{-1} \\
\Sigma _{Y|X} &=&\Sigma _{Y}-\Sigma _{XY}^{\ast }\Sigma _{X}^{-1}\Sigma _{XY}
\end{eqnarray*}

Since the distribution of$\ Y$\ conditional on $X,$ $\pi _{Y|X}\left(
y|x\right) ,$\ is also normal, 
\begin{equation*}
\partial _{xy}^{2}\log \pi _{Y|X}\left( y|x\right) =T^{\ast }\Sigma
_{Y|X}^{-1}=\Sigma _{X}^{-1}\Sigma _{XY}\left( \Sigma _{Y}-\Sigma
_{XY}^{\ast }\Sigma _{X}^{-1}\Sigma _{XY}\right) ^{-1}.
\end{equation*}%
Introducing this expression in condition (\ref{condPi}) implies%
\begin{eqnarray*}
A &=&\sigma \partial _{xy}^{2}\log \pi _{Y|X}\left( y|x\right) \\
&=&\sigma \Sigma _{X}^{-1}\Sigma _{XY}\left( \Sigma _{Y}-\Sigma _{XY}^{\ast
}\Sigma _{X}^{-1}\Sigma _{XY}\right) ^{-1}
\end{eqnarray*}%
This expression can be further simplified to%
\begin{equation*}
A=\sigma \left( \Sigma _{Y}\Sigma _{XY}^{+}\Sigma _{X}-\Sigma _{XY}^{\ast
}\right) ^{+}
\end{equation*}

These observations imply a startlingly simple way to recover the model
primitives, given the optimal matching. Intuitively, given that the observed
matching is optimal, one can regress $Y$\ on $X$. The coefficients of $X$\
constitute the corresponding entries in the matrix $T$\ and the variance of
the error term from the regression is simply the conditional variance of $Y$%
\ given $X,$\ $\Sigma _{Y|X}.$\ Then, by the equilibrium conditions, we have
that $\ \frac{A}{\sigma }=T^{\ast }\Sigma _{Y|X}^{-1}.$\ To recover the
model primitives, we have simply used the theoretical equilibrium conditions
and OLS, so we have also found a closed-form expression for the MLE for $A$\
up to a scale parameter $1/\sigma $.

As Chow and Siow \cite{CS} show, unobserved heterogeneity is crucial for
identification of matching models, even when the characteristics of the
matching populations are continuous. For example, when the affinity matrix $%
A $\ is approximately trivial, identification is not achieved if the
econometrician does not allow for unobserved heterogeneity. As in all choice
models, the model is identified up to a location and scale parameter: the
identified parameter is the rescaled affinity matrix $A/\sigma $.
Consequently, we normalize the scale of heterogeneity to one, $\sigma =1,$
and estimate the norm of the affinity matrix. Finally, if there are data on
wage transfer $\tau \left( x,y\right) $, one can recover also $B$\ and $%
\Gamma ,$\ as well as the ratio of the scale parameters $\sigma _{1}$\ \ and 
$\sigma _{2},$\ using condition (\ref{EqTransfers}) which links compensation
and the utility and profit functions.\ We formalize the identification
result for the affinity matrix $A$ in the theorem below, whose proof is
immediate given the proof of Theorem~\ref{thm:optMatching}.

\begin{theorem}
\label{thm:MatchingEstimator}Suppose that Assumptions 1 to 4 hold and let $%
\sigma =1$. Then the affinity matrix $A$\ is given by%
\begin{eqnarray}
A &=&\Sigma _{X}^{-1}\Sigma _{XY}\left( \Sigma _{Y}-\Sigma _{XY}^{\ast
}\Sigma _{X}^{-1}\Sigma _{XY}\right) ^{-1}  \label{inverseProp} \\
&=&\left( \Sigma _{Y}\Sigma _{XY}^{+}\Sigma _{X}-\Sigma _{XY}^{\ast }\right)
^{+}.  \notag
\end{eqnarray}
\end{theorem}

\subsection{Comparative Statics and Asymptotics}

In many problems, it is important to compute how an increase in the
complementarity between two characteristics in the surplus formula affects
the optimal matching. Similarly, one may want to find how the optimal
matching changes as the variances in the matching populations increase. From
an econometric point of view, the differentiation of the estimator of the
affinity matrix with respect to the summary statistics will allow us to
derive central limit theorems and to compute confidence intervals. We use
extensively matrix differentiation and the Kronecker product, for which we
now give the basic definitions (a more detailed review is given in the
Appendix). As defined in that appendix, $\mathbb{T}$ is the operator such
that 
\begin{equation*}
\mathbb{T}vec\left( M\right) =vec\left( M^{\ast }\right) .
\end{equation*}

The following theorem summarizes our results on comparative statics.

\begin{theorem}
\label{thm:Derivatives}\textit{Suppose that Assumptions 1 to 4 hold. Then:}
\end{theorem}

\textbf{(i)}\textit{\ The rate of change of the matching estimator with
respect to the covariation between the matching populations (keeping }$%
\Sigma _{X}$\textit{\ and }$\Sigma _{Y}$\textit{\ constant) is }%
\begin{equation}
\frac{\partial A}{\partial \Sigma _{XY}}=\left( A^{\ast }\otimes A\right) %
\left[ \left( \Sigma _{X}\Sigma _{XY}^{+\ast }\otimes \Sigma _{Y}\Sigma
_{XY}^{+}\right) +\mathbb{T}\right] ,  \label{cs1}
\end{equation}%
\textit{and the rates of change with respect to the variances of the
matching populations (keeping }$\Sigma _{XY}$\textit{\ constant) are}%
\begin{eqnarray}
\frac{\partial A}{\partial \Sigma _{X}} &=&-\left( A^{\ast }\otimes A\right) %
\left[ I\otimes \Sigma _{Y}\Sigma _{XY}^{+}\right] ,  \label{cs2} \\
\frac{\partial A}{\partial \Sigma _{Y}} &=&-\left( A^{\ast }\otimes A\right) %
\left[ \Sigma _{X}\Sigma _{XY}^{+\ast }\otimes I\right] .  \label{cs3}
\end{eqnarray}

\textbf{(ii) }\textit{The rate of change of the optimal cross-covariance
matrix with respect to the affinity matrix }$A$\textit{\ (keeping }$\Sigma
_{X}$\textit{\ and }$\Sigma _{Y}$\textit{\ constant) is}%
\begin{equation}
\frac{\partial \Sigma _{XY}}{\partial A}=\left[ \left( \Sigma _{X}\Sigma
_{XY}^{+\ast }\otimes \Sigma _{Y}\Sigma _{XY}^{+}\right) +\mathbb{T}\right]
^{+}\left( A^{+\ast }\otimes A^{+}\right) ,  \label{cs4}
\end{equation}%
\textit{and the rates of change with respect to the variances of the
matching populations (keeping }$A\,$\textit{constant) are }%
\begin{eqnarray}
\frac{\partial \Sigma _{XY}}{\partial \Sigma _{X}} &=&\left[ \left( \Sigma
_{X}\Sigma _{XY}^{+\ast }\otimes \Sigma _{Y}\Sigma _{XY}^{+}\right) +\mathbb{%
T}\right] ^{+}\left( I\otimes \Sigma _{Y}\Sigma _{XY}^{+}\right) ,
\label{cs5} \\
\frac{\partial \Sigma _{XY}}{\partial \Sigma _{Y}} &=&\left[ \left( \Sigma
_{X}\Sigma _{XY}^{+\ast }\otimes \Sigma _{Y}\Sigma _{XY}^{+}\right) +\mathbb{%
T}\right] ^{+}\left( \Sigma _{X}\Sigma _{XY}^{+\ast }\otimes I\right) .
\label{cs6}
\end{eqnarray}

\begin{proof}[Proof of Theorem \protect\ref{thm:Derivatives}]
We use extensively the properties of the Kronecker product and the $vec$\
operator reviewed in the Appendix and presented in much greater details by
Magnus and Neudecker \cite{MN}. The Kronecker product and the vec operator
are linked by the following formula 
\begin{equation*}
vec\left( AXB\right) =\left( B^{\ast }\otimes A\right) vec\left( X\right) 
\end{equation*}%
\ This relation is very useful in deriving the following general product
rule. Suppose that $f:R^{n}\rightarrow R^{m\times p}$\ and $%
g:R^{n}\rightarrow R^{p\times q}.$\ $\ $Then $f\cdot g:R^{n}\rightarrow
R^{m\times q}.$\ Let $I_{n}$\ denote the $n\times n$\ identity matrix. From
the relation between the Kronecker product and the vec operator 
\begin{equation*}
vec\left( I_{m}f\left( x\right) g\left( x\right) I_{q}\right) =\left( \left(
g\left( x\right) \right) ^{\ast }\otimes I_{m}\right) vec\left( f\left(
x\right) \right) =\left( I_{q}\otimes f\left( x\right) \right) vec\left(
g\left( x\right) \right) 
\end{equation*}%
Thus, we have the following product rule: 
\begin{equation}
D\left( f\left( x\right) g\left( x\right) \right) =\left( \left( g\left(
x\right) \right) ^{\ast }\otimes I_{m}\right) Df\left( x\right) +\left(
I_{q}\otimes f\left( x\right) \right) Dg\left( x\right)   \label{ProductRule}
\end{equation}%
We start with the comparative statics for $A.$\ Using the product rule (\ref%
{ProductRule}), the chain rule, and the rules of matrix differentiation
summarized in Fact 3.1, 3.2, and 3.3 of the Appendix, we find the derivative
of $\ A$\ with respect to $\Sigma _{XY}$ $:$ 
\begin{eqnarray*}
\frac{\partial A}{\partial \Sigma _{XY}} &=&-\left( A^{\ast }\otimes
A\right) \left[ -\left( \Sigma _{X}\otimes \Sigma _{Y}\right) \left( \Sigma
_{XY}^{+\ast }\otimes \Sigma _{XY}^{+}\right) -\mathbb{T}\right]  \\
&=&\left( A^{\ast }\otimes A\right) \left[ \left( \Sigma _{X}\Sigma
_{XY}^{+\ast }\otimes \Sigma _{Y}\Sigma _{XY}^{+}\right) +\mathbb{T}\right] 
\end{eqnarray*}%
Similarly, the derivatives with respect to $\Sigma _{X}$\ and $\Sigma _{Y}$\
are: 
\begin{equation*}
\frac{\partial A}{\partial \Sigma _{X}}=-\left( A^{\ast }\otimes A\right) %
\left[ I\otimes \Sigma _{Y}\Sigma _{XY}^{+}\right] 
\end{equation*}%
\begin{equation*}
\frac{\partial A}{\partial \Sigma _{Y}}=-\left( A^{\ast }\otimes A\right) %
\left[ \Sigma _{X}\Sigma _{XY}^{+\ast }\otimes I\right] 
\end{equation*}

We can also use equation (\ref{inverseProp}) to derive implicitly $\frac{%
\partial \Sigma _{XY}}{\partial A},\frac{\partial \Sigma _{XY}}{\partial
\Sigma _{X}},$\ and $\frac{\partial \Sigma _{XY}}{\partial \Sigma _{Y}}$. By
differentiation of the two sides with respect to $A$, one gets%
\begin{equation*}
-\left( A^{+\ast }\otimes A^{+}\right) =-\left( \Sigma _{X}\otimes \Sigma
_{Y}\right) \left( \Sigma _{XY}^{+\ast }\otimes \Sigma _{XY}^{+}\right) 
\frac{\partial \Sigma _{XY}}{\partial A}-\mathbb{T}\frac{\partial \Sigma
_{XY}}{\partial A}
\end{equation*}%
\begin{equation*}
\left[ \left( \Sigma _{X}\otimes \Sigma _{Y}\right) \left( \Sigma
_{XY}^{+\ast }\otimes \Sigma _{XY}^{+}\right) +\mathbb{T}\right] \frac{%
\partial \Sigma _{XY}}{\partial A}=\left( A^{+\ast }\otimes A^{+}\right)
\end{equation*}%
Solving for $\frac{\partial \Sigma _{XY}}{\partial A},$ we obtain 
\begin{eqnarray*}
\frac{\partial \Sigma _{XY}}{\partial A} &=&\left[ \left( \Sigma _{X}\otimes
\Sigma _{Y}\right) \left( \Sigma _{XY}^{+\ast }\otimes \Sigma
_{XY}^{+}\right) +\mathbb{T}\right] ^{+}\left( A^{+\ast }\otimes A^{+}\right)
\\
&=&\left[ \left( \Sigma _{X}\Sigma _{XY}^{+\ast }\otimes \Sigma _{Y}\Sigma
_{XY}^{+}\right) +\mathbb{T}\right] ^{-1}\left( A^{+\ast }\otimes
A^{+}\right)
\end{eqnarray*}%
Similarly, by differentiation with respect to $\Sigma _{X}$\ and $\Sigma
_{Y},$\ one gets 
\begin{eqnarray*}
\frac{\partial \Sigma _{XY}}{\partial \Sigma _{X}} &=&\left[ \mathbb{T}%
+\left( \Sigma _{X}\otimes I\right) \left( I\otimes \Sigma _{Y}\right)
\left( \Sigma _{XY}^{+\ast }\otimes \Sigma _{XY}^{+}\right) \right]
^{+}\left( I\otimes \Sigma _{Y}\Sigma _{XY}^{+}\right) \\
&=&\left[ \left( \Sigma _{X}\Sigma _{XY}^{+\ast }\otimes \Sigma _{Y}\Sigma
_{XY}^{+}\right) +\mathbb{T}\right] ^{+}\left( I\otimes \Sigma _{Y}\Sigma
_{XY}^{+}\right) \\
\frac{\partial \Sigma _{XY}}{\partial \Sigma _{Y}} &=&\left[ \left( I\otimes
\Sigma _{Y}\right) \left( \Sigma _{X}\otimes I\right) \left( \Sigma
_{XY}^{+\ast }\otimes \Sigma _{XY}^{+}\right) +\mathbb{T}\right] ^{+}\left(
\Sigma _{X}\Sigma _{XY}^{+\ast }\otimes I\right) \\
&=&\left[ \left( \Sigma _{X}\Sigma _{XY}^{+\ast }\otimes \Sigma _{Y}\Sigma
_{XY}^{+}\right) +\mathbb{T}\right] ^{+}\left( \Sigma _{X}\Sigma
_{XY}^{+\ast }\otimes I\right) .
\end{eqnarray*}%
\bigskip
\end{proof}

Theorem \ref{thm:Derivatives}, part i presents the comparative statics of
the solution of the identification problem as the observed cross-covariance
matrix varies. In addition, we explore how the estimated surplus technology
changes as the characteristics of the matching populations change. These
formulas can be use to evaluate the local stability of the estimated surplus
technology$.$\ In particular, given the formulas, one can also easily
totally differentiate a change in the surplus technology into a component
due to change in $X,$\ a component due to change in $Y,$\ and a component
due to change in $\Sigma _{XY}.$ It can also be used to derive asymptotic
properties, using a standard delta method and the expression of the
derivatives (\ref{cs1})-(\ref{cs3}). To analyze the precision of our
estimates, one can apply the closed-form formulas again to the asymptotic
results in Theorem 2 and Corollary 1 of Dupuy and Galichon \cite{DG}.

Theorem \ref{thm:Derivatives}, part ii explores how the solution of the
equilibrium problem varies with a change in the surplus technology $A,$\ and
the characteristics of the two sides in the market. Without closed-forms,
one can study the effect of a change in the surplus technology or the
characteristics of the matching populations only numerically which has the
disadvantage of being computationally intensive and often difficult to
interpret the causes of the changes. In contrast within our model, we can
derive exactly how the solutions are going to evolve as the structural
parameters change, which of both applied and empirical interest. Similarly
to above, we can decompose the overall change in the equilibrium matching
into a component due to changes in the matching technology and components
due to changes in the characteristics of the matching populations $X$\ and $%
Y.$\ As a consequence, this theorem can be used to easily evaluate the
impact of counterfactual experiments.

Finally, one can test the quadratic-normal setting through an easy to
construct overidentification test. Given the observed match data, the
econometrician can find an estimate of the affinity matrix, $\widehat{A},$\
and then given $\widehat{A}$\ and the marginals predict the optimal matching 
$\widehat{\pi }_{XY}.$\ Under the hypothesis that the normal-quadratic
setting is correct, the distance between $\widehat{\pi }_{XY}$\ and the
empirically observed $\pi _{XY}$\ should be small, where the formal test
statistics have asymptotic distributions that can be easily found given the
comparative statics from above.

\section{Conclusion\label{sec:conclusion}}

This paper adopts the framework of Choo and Siow \cite{CS} and Galichon and
Salani\'{e} \cite{GS} and imposes additional functional form and
distributional assumptions that allow us to derive closed-form formulas for
the surplus and the optimal matching. In this setting, one may solve easily
problems that in the past have been computationally intensive. Moreover, the
results allow for the characterization of the optimal solution and how it
changes as the primitives of the model change. The model in this paper is
suited to static matching problems with exogenously fixed populations in
which all participants are matched. Thus, it covers a number of contexts
with potential for future applied work, such as matching of workers with
tasks, scheduling, matching between specialists and firms or top managers
and firms.

While our framework is Transferable Utility (TU), this paper's main argument
may be carried without difficulty into some instances of the Nontransferable
Utility (NTU) framework. Indeed, a recent contribution by Menzel (2014)
implies that in the case with no singles and a particular structure of
(logit) heterogeneity, the equilibrium matching in the TU and the NTU cases
coincide. Hence, the results of the present paper would apply to that
framework as well.

\appendix

\section*{Appendix: Matrix Differentiation}

The Kronecker product and the vectorization operation are extremely useful
when it comes to studying asymptotic properties involving matrices. The idea
is that matrices $m\times n$, where $m$\ stands for the number of rows and $%
n $\ for the number of $\ $columns, can be seen as $mn\times 1$\ vectors in $%
R^{mn}$, and linear operation on such matrices can be seen as higher order
matrices. To do this, the fundamental tool is the vectorization operation,
which vectorizes a matrix by stacking its columns. Introduce $\tau _{mn}$\ a
collection of invertible maps from $\left\{ 1,...,m\right\} \times \left\{
1,...,n\right\} $\ onto $\left\{ 1,...,mn\right\} $, such that $\tau
_{mn}\left( i,j\right) =m\left( j-1\right) +i$.

\begin{definition}
For $\left( M\right) $ a $m\times n$ matrix, $vec\left( M\right) $ is the
vector $v\in \mathbb{R}^{mn}$ such that $v_{\tau _{mn}\left( i,j\right)
}=M_{ij}$.
\end{definition}

\bigskip

Next, we introduce the transposition tensor $\mathbb{T}_{m,n}$ as the $%
mn\times mn$ matrix such that%
\begin{equation*}
\mathbb{T}_{m,n}vec\left( M\right) =vec\left( M^{\ast }\right) .
\end{equation*}

The matrix operator $T_{m,n}$\ is a permutation matrix with zeros and a
single 1 on each row and column. Note that $\mathbb{T}_{m,n}=\mathbb{T}%
_{n,m}^{-1}$, so $\mathbb{T}_{m,n}\mathbb{T}_{n,m}vec\left( M\right)
=vec\left( M\right) .$ Furthermore, $\mathbb{T}_{m,n}=\mathbb{T}_{n,m}^{\ast
}$. The next definition deals with Kronecker product, which is closely
related to vectorization.

\begin{definition}
Let $A$\ be a $m\times p$\ matrix and $B$\ an $n\times q$\ matrix. One
defines the Kronecker product $A\otimes B$\ as the $mn\times pq$\ matrix
such that%
\begin{equation*}
\left( A\otimes B\right) _{n\left( i-1\right) +k,q\left( j-1\right)
+l}=A_{ij}B_{kl}.
\end{equation*}
\end{definition}

The following fundamental property characterizes the Kronecker product.

\begin{fact}
For all $q\times p$\ matrix $X$,%
\begin{equation*}
vec\left( BXA^{T}\right) =\left( A\otimes B\right) vec\left( X\right)
\end{equation*}
\end{fact}

The following important basic properties follow.

\begin{fact}
Let $A$ be a $m\times p$ matrix and $B$ an $n\times q$ matrix. Then:

1. (Associativity) $\left( A\otimes B\right) \otimes C=A\otimes \left(
B\otimes C\right) .$

2. (Distributivity) $A\otimes \left( B+C\right) =A\otimes B+A\otimes C.$

3. (Multilinearity) For $\lambda $ and $\mu $ scalars, $\lambda A\otimes \mu
B=\lambda \mu \left( A\otimes B\right) $

4. For matrices of appropriate size, $\left( A\otimes B\right) \left(
C\otimes D\right) =\left( AC\right) \otimes \left( BD\right) $.

5. $\left( A\otimes B\right) ^{\ast }=A^{\ast }\otimes B^{\ast }$.

6. If $A$ and $B$ are invertible, $\left( A\otimes B\right)
^{-1}=A^{-1}\otimes B^{-1}$.

7. For vectors $a$ and $b$, $a^{\prime }\otimes b=ba^{\prime }$ (in
particular, $aa^{\prime }=a^{\prime }\otimes b$).

8. If $A$ and $B$ are square matrices of respective size $m$ and $n$, 
\begin{equation*}
\det \left( A\otimes B\right) =\left( \det A\right) ^{m}\left( \det B\right)
^{n}.
\end{equation*}

9. $Tr\left( A\otimes B\right) =Tr\left( A\right) Tr\left( B\right) $.

10. $rank\left( A\otimes B\right) =rank\left( A\right) rank\left( B\right) $.

11. The singular values of $A\otimes B$ are the product of the singular
values of $A$ and those of $B$.
\end{fact}

Let $f$ be a smooth map from the space of $m\times p$ matrices to the space
of $n\times q$ matrix. Define $\frac{df\left( A\right) }{dA}$ as the $\left(
nq\right) \times \left( mp\right) $ matrix such that for an $m\times p$
matrix $X$,%
\begin{equation*}
vec\left( \lim_{e\rightarrow 0}\frac{f\left( A+eX\right) -f\left( A\right) }{%
e}\right) =\frac{df\left( A\right) }{dA}.vec\left( X\right) .
\end{equation*}

We use the notation $A^{-\ast }$ for $\left( A^{\ast }\right) ^{-1}$.

\begin{fact}
Let $A$ be a $m\times p$ matrix and $B$ an $n\times q$ matrix. Then:

1. $\frac{d\left( AXB\right) }{dX}=B^{\ast }\otimes A$.

2. $\frac{dA^{\ast }}{dA}=\mathbb{T}_{m,p}$.

3. $\frac{dA^{-1}}{dA}=-\left( A^{-\ast }\otimes A^{-1}\right) $.

4. $\frac{dA^{2}}{dA}=I\otimes A+A^{T}\otimes I$

5. For $A$ symmetric, $\frac{dA^{1/2}}{dA}=\left( I\otimes
A^{1/2}+A^{1/2}\otimes I\right) ^{-1}$.
\end{fact}

{\small $^{\S }$}\textit{\ Department of Economics, \'{E}cole polytechnique.
Address: Route de Saclay, 91128 Palaiseau, France. E-mail:
raicho.bojilov@polytechnique.edu.} 

\textit{$^{\dag }$ Economics Department and Courant Institute, New York
University and Economics Department, Sciences Po. Address: Department of
Economics. 19 W 4th Street, New York, NY 10012, USA. Email: ag133@nyu.edu. } 


\begin{thebibliography}{99}
\bibitem{AP} Angeletos, G.-M., Pavan, A. (2007). "Efficient Use of
Information and Social Value of Information." \emph{Econometrica} 75(4),
1103--1142.

\bibitem{B} Becker, G. (1973). "A Theory of Marriage, Part I," \emph{Journal
of Political Economy} 81, 813-846.

\bibitem{BR} Brown, J. N., and Rosen, H. (1982). \textquotedblleft On the
Estimation of Structural Hedonic Price Models.\textquotedblright\ \emph{%
Econometrica} 50, 765--68.

\bibitem{BW} Ben-Akiva, M. and Watanatada, T. (1981). \textquotedblleft
Application of a Continuous Spatial Choice Logit Model.\textquotedblright\
In \emph{Structural Analysis of Discrete Choice Data with Econometric
Applications} (C.F. Manski and D. McFadden, eds.), MIT Press.

\bibitem{BenIsrael} Ben-Israel, Adi; Thomas N.E. Greville (2003). \textit{%
Generalized Inverses}. Springer-Verlag.

\bibitem{CMN} Chiappori, P.A., McCann, R., Nesheim, L. (2010). "Hedonic
Price Equilibria, Stable Matching, and Optimal Transport: Equivalence,
Topology, and Uniqueness," \emph{Economic Theory} 42(2), 317-354.

\bibitem{COQ} Chiappori, P.-A., Oreffice, S. and Quintana-Domeque, C.
(2012). \textquotedblleft Fatter Attraction: Anthropometric and
Socioeconomic Matching on the Marriage Market,\textquotedblright\ \emph{%
Journal of Political Economy} 120 (4), 659--695.

\bibitem{CSW} Chiappori, P.-A., Salani\'{e}, B., and Weiss, Y. (2008).
"Assortative Matching on the Marriage Market: a Structural Investigation,"
working paper.

\bibitem{CS} Choo, E., and Siow, E. (2006). \textquotedblleft Who Marries
Whom and Why\textquotedblright .\ \emph{Journal of Political Economy} 114,
172--201.

\bibitem{Cosslett} Cosslett, S. (1988). \textquotedblleft Extreme-Value
Stochastic Processes: A Model of Random Utility Maximization for a
Continuous Choice Set,\textquotedblright\ Technical report, Ohio State
University.

\bibitem{Dagsvik} Dagsvik, J. (1994). \textquotedblleft Discrete and
Continuous Choice, Max-Stable Processes, and Independence from Irrelevant
Attributes,\textquotedblright\ \emph{Econometrica} 62, 1179--1205.

\bibitem{DLMS} Decker, C., Lieb, E., McCann, R. and Stephens, B. (2013).
\textquotedblleft Unique Equilibria and Substitution Effects in a Stochastic
Model of the Marriage Market\textquotedblright . \emph{Journal of Economic
Theory} 148, 778--792.

\bibitem{DL} Dowson, D.C., and B.V. Landau (1982). \textquotedblleft The Fr%
\'{e}chet Distance Between Multivariate Normal
Distributions\textquotedblright . \emph{Journal of Multivariate Analysis}
12, 450--455.

\bibitem{DG} Dupuy, A., and Galichon, A. (2014). \textquotedblleft
Personality Traits and the Marriage Market\textquotedblright . Forthcoming
in the \emph{Journal of Political Economy}.

\bibitem{E} Ekeland, I. (2010). "Existence, Uniqueness and Efficiency of
Equilibrium in Hedonic Markets with Multidimensional Types"\textit{\ \emph{%
Economic Theory} }42, p. 275--315.

\bibitem{EGH} Ekeland, I., Galichon, A., Henry, M. (2010) "Optimal
Transportation and the Falsifiability of Incompletely Specified Economic
Models", \emph{Economic Theory} 42, p. 355 - 374

\bibitem{EHN} Ekeland, I., J. J. Heckman, and L. Nesheim (2004).
"Identification and Estimation of Hedonic Models." \textit{Journal of
Political Economy }112(S1), S60-S109. Paper in Honor of Sherwin Rosen: A
Supplement to Volume 112.

\bibitem{F} Fox, J. (2010). \textquotedblleft Identification in Matching
Games\textquotedblright . \emph{Quantitative Economics} 1(2), pp. 203--254.

\bibitem{FK} Fox, J. (2014). \textquotedblleft Estimating Matching Games
with Transfers\textquotedblright . Manuscript.

\bibitem{GL} Gabaix, X. and Landier (2008). \textquotedblleft Why Has CEO
Pay Increased So Much?\textquotedblright . \emph{Quaterly Journal of
Economics} 123(1), p. 49--100.

\bibitem{GS} Galichon, A., and Salani\'{e}, B. (2010). \textquotedblleft
Matching with Trade-offs: Revealed Preferences over Competing
Characteristics\textquotedblright . Technical report.

\bibitem{GS2} Galichon, A., and Salani\'{e}, B. (2014). \textquotedblleft
Cupid's Invisible Hand: Social Surplus and Identification in Matching
Models\textquotedblright . Working paper.

\bibitem{GP} Gomes, R., Pavan, A. (2015). "Many-to-Many Matching and Price
Discrimination". Working paper.

\bibitem{Graham} Graham, B. (2011). \textquotedblleft Econometric Methods
for the Analysis of Assignment Problems in the Presence of Complementarity
and Social Spillovers," in \emph{Handbook of Social Economics}, ed. by J.
Benhabib, A. Bisin, and M. Jackson. Elsevier.

\bibitem{GOZ} Gretsky, N., Ostroy, J., Zame, W., (1992). "The Nonatomic
Assignment Model," \emph{Economic Theory}, vol. 2(1), p. 103-27.

\bibitem{HHJK} Hsieh, C.-T., Hurst, E., Jones, C., Klenow, P. (2013).
\textquotedblleft The Allocation of Talent and US Economic
Growth.\textquotedblright\ \emph{NBER Working Paper} No. 18693.

\bibitem{L} Lindenlaub I. \ (2013). \textquotedblleft Sorting
Multidimensional Types: Theory and Application\textquotedblright\ Manuscript.

\bibitem{Machado} de Caralho Montalvao Machado, J. (2013). "Essays on
Competition in Bipartite Matching and in Policy Combinations." Dissertation.

\bibitem{MN} Magnus, J. and Neudecker H. (2007). $\emph{M}$\emph{atrix
Differential Calculus with Applications in Statistics and Econometrics}, 3rd
Edition. John Wiley and Sons Ltd. \textbf{\ }

\bibitem{Menzel} Menzel, K. (2014). \textquotedblleft Large Matching Markets
as Two-Sided Demand Systems.\textquotedblright\ Manuscript.

\bibitem{OP} Olkin, I. and F. Pukelsheim (1982). \textquotedblleft The
Distance between Two Random Vectors with Given Dispersion
Matrices.\textquotedblright\ \emph{Linear Algebra Appl.} 48, 257-263.

\bibitem{RR} Resnick, S., and R. Roy (1991). \textquotedblleft Random USC
functions, Max-Stable Processes and Continuous Choice,\textquotedblright\ 
\emph{Annals of Applied Probability} 1(2), 267--292.

\bibitem{Rs} Rosen, S. (1974). \textquotedblleft Hedonic Prices and Implicit
Markets: Product Differentiation in Pure Competition.\textquotedblright\ 
\emph{Journal of Political Economy} 82, 34--55.

\bibitem{R} Roy, A. D. (1951). \textquotedblleft Some Thoughts on the
Distribution of Earnings,\textquotedblright\ \emph{Oxford Economic Papers} 3
(2), 135--146.

\bibitem{SS} Shimer, R. and Smith L. (2000). "Assortative Matching and
Search," \emph{Econometrica} 68(2), 343-369.

\bibitem{Tj} Tinbergen, J. (1956). "On the Theory of Income Distribution." 
\emph{Weltwirtschaftliches Archiv}, 155-175.

\bibitem{T} Tervio, M. (2008). \textquotedblleft The difference that CEO\
make: An Assignment Model Approach,\textquotedblright\ \emph{American
Economic Review} 98 (3), pp. 642--668.

\bibitem{Vi} Villani, C. (2004). \emph{Topics in Optimal Transportation}.
AMS.
\end{thebibliography}
\end{document}